\documentclass[a4paper,10pt]{article}
\usepackage{graphicx}
\usepackage{caption}

\usepackage{amsmath}
\newcommand{\la}{\label}
\newcommand{\p}{\partial}
\newcommand{\be}{\begin{equation}}
\newcommand{\ee}{\end{equation}}
\newtheorem{thm}{Theorem}[section]
\newcommand{\ba}{\begin{eqnarray}}
\newcommand{\ea}{\end{eqnarray}}

\begin{document}
\centerline{\bf \begin{Large} The Penrose inequality for perturbations \end{Large}} 

\centerline{\bf \begin{Large} of the Schwarzschild initial data \end{Large}}                                                                                    

\bigskip

\centerline{J. Kopi\'nski and J. Tafel}

\centerline{Faculty of Physics, University of Warsaw,}
\centerline{Pasteura 5, 02-093 Warsaw, Poland}
\begin{abstract}
We show  that in the conformally flat case 
the Penrose inequality  is satisfied for the Schwarzschild initial data with a small addition of the  axially symmetric traceless exterior curvature. In this class the inequality is saturated only for data related to special sections of  the Schwarzschild spacetime.
\end{abstract}

\null

\section{Introduction}

The mass $M$ and the surface area $|S_h|$ of the event horizon in the Kerr metric satisfy the  inequality
\be\la{1a}
M\geq \sqrt{\frac{|S_h|}{16\pi}},
\ee
saturated by the Schwarzschild solution. Arguments based on the singularity theorems, no-hair theorems and thermodynamics of black holes led Penrose to the cosmic censorship conjecture and hypothesis that inequality (\ref{1a}) should be  satisfied in physically realistic spacetimes  on any initial surface $S$ with an internal horizon \cite{p}. Simultaneously  Geroch \cite{g}, in his search for a proof of the positive energy theorem, proposed a definition of a quasilocal mass and showed its  monotonicity   under the inverse mean curvature flow (IMCF). This method was then pointed  out by Jang and Wald \cite{jw} as a proper tool  to prove (\ref{1a}) provided mathematical problems with singularities of IMCF would be resolved. Two decades later this approach was completed  by  Huisken and Ilmanen \cite{hi} under the assumption that  the Ricci scalar of   $S$ is  positive and the internal horizon is  a minimal surface.  These conditions are natural if the initial data on $S$ are time symmetric. In more general setting the  exterior curvature of $S$ should be admitted   and the horizon would  be  a kind of a trapped surface. Some results in this direction
were obtained by Malec, Mars and Simon \cite{mms} (see also  \cite{rm}) who applied  the Geroch method  to the Hawking quasilocal mass.  An extended  discussion of results and ideas on the Penrose inequality  can be found in a review by Mars \cite{m}.

In this paper we present an approach  to the Penrose inequality based on the conformal method of solving  constraints  in general relativity \cite{l,y}.
Initial data for the vacuum  Einstein equations consists of a 3-dimensional Riemannian metric $g'_{ij}$ and a symmetric tensor $K'_{ij}$ constrained by  the equations
\be\la{2}
\nabla'{}_i\big(K'^i{}_j-H'\delta^i{}_j\big)=0\ ,
\ee
\be\la{3h}
  R'+H'^2-K'^2=0\ ,
\ee 
where $H'=K'{}_i{}^i$ and $K'^2=K'{}_{ij}K'^{ij}$.
Let $S$ contains a 2-dimensional closed compact surface $S_h$, which is  marginally outer trapped (MOTS),
\be\la{3j}
H'-K'_{nn}+ 2h'=0\ .
\ee
Here $K'_{nn}=n'^in'^jK'{}_{ij}$,  $n'^i$ is the unit outer normal vector to $S_h$ and  $h'=\frac 12\nabla'{}_i n'^i$ is the mean curvature of $S_h$.
In order to find initial data admitting MOTS we start with preliminary data  $g_{ij},K_{ij}$  satisfying
\be\la{4c}
H=0
\ee
and
\be\la{4b}
\nabla_iK^i_{\ j}=0\ .
\ee
Then data
\begin{equation}\la{6a}
g'_{ij}=\psi^4 g_{ij},\quad K'^i_{\ j}=\psi^{-6} K^i_{\ j}\ ,\ \ \psi>0
\end{equation}
also satisfy  (\ref{4c})-(\ref{4b}) and the Hamiltonian constraint (\ref{3h}) is equivalent to the 
 Lichnerowicz equation 
\begin{equation}\la{7a}
\bigtriangleup\psi=\frac 18 R\psi-\frac 18  K^2\psi^{-7}\ ,
\end{equation}
where $\bigtriangleup$ and $R$ are, respectively, the Laplace operator and the Ricci scalar related to $g_{ij}$.
If $\psi$ satisfies in addition the boundary condition
\be\la{8f}
n^i\p_i \psi+\frac 12 h\psi-\frac 14 K_{nn} \psi^{-3}=0\ \ \mathrm{on}\ \  S_h
\ee
then the surface $S_h$ becomes MOTS with respect to the primed data. The existence theorems for equations (\ref{7a})-(\ref{8f}) were proved by Dain \cite{d} and Maxwell \cite{ma,ma1} (see \cite{dhkm} for a review of results on solvability of the Lichnerowicz equation in different settings). 

It is known that for some asymptotically flat data with MOTS  inequality (\ref{1a}) is not  true \cite{bd}.  A more plausible inequality follows if  $|S_h|$ is replaced by a minimal area of surfaces enclosing $S_h$,
\be\la{9c}
M\geq \sqrt{\frac{A_{min}(S_h)}{16\pi}}\ .
\ee
Moreover, as $S_h$ one can take  an outermost MOTS
\be\la{9d}
M\geq \sqrt{\frac{A_{min}(S_{out})}{16\pi}}\ .
\ee
In the case of solutions to equations (\ref{7a})-(\ref{8f}) we are not able to control if the resulting  MOTS is outermost and if there are surfaces outside $S_h$ with a smaller area. Still, for a class of data we can prove  (\ref{1a}) with just area of the MOTS. This inequality is stronger than (\ref{9c}) but may  be weaker than (\ref{9d}). In fact we prove the stronger version of the Penrose inequality 
\be\la{8k}
E^2-p^2\geq \frac{|S_h|}{16\pi}+\frac{4 \pi}{|S_h|}J^2\ ,
\ee
which makes sense in the case of axial symmetry. Here  $E,p,J$ are, respectively,  the total energy, momentum and angular momentum of data.

 In what follows we assume that data are  axially symmetric and initial metric is conformally flat. In order to estimate components of (\ref{8k})   we  study equations (\ref{7a})-(\ref{8f}) perturbatively assuming expansions of $\psi$ into powers of a small parameter proportional to the preliminary exterior curvature.
Our main conclusion is that in the conformally flat case 
the Penrose inequality (\ref{8k}) is preserved under a small addition of the  axially symmetric traceless exterior curvature to the Schwarzschild initial data. Moreover, within this class inequality  (\ref{8k}) is saturated only for 1-parameter spherically symmetric data which correspond to special slices of the Schwarzschild spacetime. The main value of our results is that, unlike in most approaches based on the monotonicity of the Geroch mass (except \cite{mms}), we do not assume that the internal horizon is a minimal surface. 

The considered class of metrics cannot contain the Kerr metric since the latter does not admit a conformally flat slice. Still, our data admit nontrivial angular momentum $J$ which cannot be radiated away according to analysis in \cite{fm}. Thus, if metric tends to a stationary state without a naked singularity its end state should be just the Kerr metric with angular momentum $J$.  Including the Kerr metric into our considerations would require a reformulation of our approach in order to avoid its dependence on exact solutions of the momentum constraint  and on expansions into the spherical harmonics. An  easier generalization seems to be admitting nonmaximal data ($H\neq 0$), perhaps without axial symmetry, while still keeping the conformal flatness. 

In section 2 we find convenient expressions for the total energy $E$ and the horizon area $|S_h|$ if the preliminary metric is flat and the boundary $S_h$ is the 2-dimensional sphere. In section 3 we apply these formulas to axially symmetric solutions of the momentum constraint. We expand  the exterior curvature into the Legendre polynomials and find the minimal value of the quantity $P_I=E^2-|S_h|/16\pi$ for a fixed value of $K_{nn}$ on  $S_h$. It turns out that the Penrose inequality (\ref{8k}) is satisfied in the second order in $K_{ij}$ for all considered data except the generalized Bowen-York data \cite{by,beig}. The latter need fourth order calculation which is done in section 4. Again, the Penrose inequality follows. This part is illustrated by a numerical simulation for a 2-parameter family of data.

\section{A change of total energy and the horizon area under the conformal transformation}
The construction of initial data via the conformal method with $H=0$ is relatively simple if the  initial metric $g_{ij}$ is  flat. Then one can 
solve explicitly the momentum constraint  and express the exterior curvature in terms of arbitrary functions (see \cite{cm,c} for the axially symmetric case and \cite{t} for general one).  

We assume that $g_{ij}=\delta_{ij}$ and the initial surface is given by  $S=R^3\backslash B(0,\frac m2)$, where  $B(0,\frac m2)$ is an open ball  with radius $m/2$ and the spherical boundary  $S_h$. If $K_{ij}=0$  then solution of (\ref{7a}) and (\ref{8f}) reads
\be\la{9}\psi=\psi_0=1+\frac{m}{2r}
\ee
and transformation (\ref{6a}) leads  to the Schwarzschild initial metric  with mass $m$ and  $S_h$ being a minimal surface. In this case the Penrose inequality is saturated. Our goal is to investigate  how this inequality is affected by addition of  a small amount of the exterior curvature to these data.

Let $K_{ij}$ be a traceless solution  of the momentum constraint which is asymptotically flat in the sense that
\be\la{9a}
K_{ij}=O(r^{-2})\ ,\ \ \ \p_pK_{ij}=O(r^{-3})\ \ \mathrm{if}\ \ r\longrightarrow\infty\ .
\ee
Then $K_{ij}$ belongs to the weighted Sobolev space $W^{1,2}_{\rho-1}$, where $-1<\rho<0$. Now we can use results of Maxwell \cite{ma,ma1} to show that equations 
(\ref{7a})-(\ref{8f}) admit a solution $\psi$. From Proposition 3 in \cite{ma} and  existence of transformation (\ref{8f}) to the Schwarzschild data one obtains that the Yamabe constant is positive, $\lambda_g>0$ (this property follows also from the Sobolev inequality on $S$ proved in Theorem 6.2 in \cite{tj}). Now, existence of $(\psi-1)\in W^{2,2}_{\rho}$ for $K_{nn}|_{S_h}\leq 0$ is assured by Theorem 4.2 in \cite{ma1}  and existence of $\psi$ for $2h\geq K_{nn}|_{S_h}\geq 0$ follows from Theorem 1 in \cite{ma} (note that we use definitions of $K_{ij}$ and $h$ with opposite signs). The upper limit in the latter case is not relevant for us since we will assume smallness of $K^2$. There are no existence theorems for $K_{nn}$ of indefinite sign on $S_h$. 

In what follows we assume that $\psi$ exists and  
\be\la{9b}
\psi\longrightarrow 1\ \ \mathrm{if}\ \ r\longrightarrow\infty\ .
\ee
In order to extract an information about energy from (\ref{7a})-(\ref{8f}) it is convenient to use  tilded variables defined by
\begin{equation}\label{tild_var}
\tilde \psi=\frac{\psi}{\psi_0}\ ,\ \ \tilde g_{ij}=\psi_0^4 g_{ij},\quad \tilde K_{ij}=\psi_0^{-2} K_{ij}\ .
\end{equation}
Equations (\ref{7a})-(\ref{8f}) are equivalent to
\begin{equation}\label{20}
\tilde\bigtriangleup\tilde\psi=-\frac 18  \tilde K^2\tilde\psi^{-7}\ ,
\end{equation}
\be\label{21a}
\p_r\tilde \psi=\frac {1}{16} \tilde K_{rr} \tilde\psi^{-3}\ \ at\ r=m/2\ ,
\ee
where $\tilde \bigtriangleup$ is the Laplace operator corresponding to the Schwarzschild metric $\tilde g$.
The ADM formula for the total energy of  the ultimate data (\ref{6a}) yields
\be\la{12d}
E=m-\frac{1}{2\pi}\lim_{r\rightarrow\infty}\oint_{S_r}{r^2\tilde\psi_{,r}d\sigma}\ ,
\ee
where $d\sigma$ is the surface measure on the unit sphere. Integrating (\ref{20}) over $S$ with the volume element $\tilde\eta$ defined by  $\tilde  g$ one obtains
\be\la{13}
E=m+\frac{1}{16\pi}\int_S{\tilde K^2\tilde\psi^{-7}\tilde\eta}-\frac{m^2}{32\pi}\oint_{S_h}{\tilde K_{rr}\tilde\psi^{-3}d\sigma}\ .
\ee
Note that energy $E$ is finite thanks to conditions (\ref{9a}) and (\ref{9b}).

Let  $K_{ij}$ be proportional to a small parameter $\epsilon$. We assume that  $\psi$ can be expanded into the sum
\be\la{10}
\psi=\psi_0+\psi_1+\psi_2+...
\ee
where terms $\psi_n$ are of the order $\epsilon^n$. Hence
\be\la{10a}
\tilde\psi=1+\tilde\psi_1+\tilde\psi_2+...
\ee
and, up to $\epsilon^2$, equation (\ref{13}) reads
\be\la{13c}
E=m+\frac{1}{16\pi}\int_S{\tilde K^2\tilde\eta}-\frac{m^2}{32\pi}\oint_{S_h}{\tilde K_{rr}(1-3\tilde\psi_1)d\sigma}\ .
\ee
In terms of untilded quantities formula (\ref{13c}) takes the form
\be\la{14}
E=m+\frac{1}{16\pi}\int_S{K^2\psi_0^{-6}\eta}-\frac{m^2}{128\pi}\oint_{S_h}{K_{rr}(1-\frac 32\psi_1)d\sigma}\ ,
\ee
where $\eta$ is the flat volume element.
An advantage of expression (\ref{14}) over (\ref{12d}) is that to calculate its r. h. s. we don't need $\psi_2$. It is sufficient to find $\psi_1$  satisfying the flat Laplace equation
\be\la{14a}
\bigtriangleup\psi_1=0
\end{equation}
and the boundary conditions
\be\label{15}
\p_r\psi_1+\frac 1m\psi_1=\frac {1}{32} K_{rr}\ \ \mathrm{at}\ \ r=m/2\ ,
\ee
\be\la{16}
\psi_1= 0\ \ \mathrm{at}\ \ r=\infty\ .
\ee

Up to $\epsilon^2$ the ultimate surface area of $S_h$ is
\be\la{26}
|S_h|=\oint_{S_h}{\psi^4r^2d\sigma}=4m^2\oint_{S_h}{(1+4\tilde\psi_1+4\tilde\psi_2+6\tilde\psi_1^2)d\sigma}\ .
\ee
In order to eliminate $\tilde\psi_2$ from (\ref{26}) let us integrate  an approximate version of equation (\ref{20}) over  spherical coordinates. Hence
\be\la{28f}
r^{-2}\psi_0^{-6}(r^2\psi_0^2 \langle \tilde\psi \rangle_{,r})_{,r}=-\frac 18\langle \tilde K^2\rangle\ ,
\ee
where 
\be\la{27}
\langle f\rangle =\oint_{S_r}{fd\sigma}
\ee
for any function $f$.
Twice integrating (\ref{28f}) over $r$ yields
\be\la{29}
\langle \tilde\psi\rangle=\frac 18\int_r^{\infty}{\frac{dr'}{r'^2\psi_0^2}\int_{\frac m2}^{r'}{\frac{\tilde r^2\langle K^2\rangle}{\psi_0^6}d\tilde r}}-\frac{c_1}{r\psi_0}+c_2\ ,
\ee
where $c_1$ and $c_2$ are constants.
 Constant $c_1$ follows from integration of (\ref{21a}) 
\be\la{31}
c_1=\frac{m^2}{64}\langle K_{rr}(1-\frac 32\psi_1)\rangle_h\ ,
\ee
where $\langle \rangle_{h}$ denotes the integral (\ref{27}) for $r=\frac m2$.
 Value
$$c_2=4\pi$$ follows from the asymptotic condition $\psi\longrightarrow 1$.

Substituting (\ref{29})  and (\ref{31}) into (\ref{26}) leads to
\be\la{32}
|S_h|=16\pi m^2+6m^2\langle\psi_1^2\rangle_{h}-\frac {m^3}{4}\langle K_{rr}(1-\frac 32\psi_1)\rangle_{h}+2 m^2\int_{\frac m2}^{\infty}{\frac{dr'}{r'^2\psi_0^2}\int_{\frac m2}^{r'}{\frac{r^2\langle K^2\rangle}{\psi_0^6}dr}}\ .
\ee
The double integral in (\ref{32})  simplifies if we change the order of integration 
(here $\theta$ is the Heaviside function)
\ba\la{32a}
\int_{\frac m2}^{\infty}\frac{dr'}{r'^2\psi_0^2}\int_{\frac m2}^{\infty}\theta (r'-r)\frac{r^2\langle K^2\rangle}{\psi_0^6}dr=\int_{\frac m2}^{\infty}dr\frac{r^2\langle K^2\rangle}{\psi_0^6}\int_{ r}^{\infty}\frac{dr'}{r'^2\psi_0^2}\ .
\ea
Due to (\ref{32a}) formula (\ref{32}) takes the form
\be\la{32f}
|S_h|=16\pi m^2+6m^2\langle\psi_1^2\rangle_{h}-\frac {m^3}{4}\langle K_{rr}(1-\frac 32\psi_1)\rangle_{h}+2 m^2\int_{\frac m2}^{\infty}\frac{r\langle K^2\rangle}{\psi_0^7}dr\ .
\ee

With use of  notation (\ref{27}) expression (\ref{14}) reads
\be\la{13a}
E=m-\frac{m^2}{128\pi}\langle K_{rr}(1-\frac 32\psi_1)\rangle_{h}+\frac{1}{16\pi}\int_{\frac m2}^{\infty}\frac{r^2\langle K^2\rangle}{\psi_0^6}dr\ .\ee
From (\ref{32f}) and (\ref{13a}) one obtains 
\be\la{33b}
E^2-\frac{|S_h|}{16\pi}=P_I\ ,
\ee
where
\be\la{32b}
P_I=(\frac{m^2}{128\pi})^2\langle K_{rr}\rangle_{h}^2-\frac{3m^2}{8\pi}\langle \psi_1^2\rangle_{h}+\frac{m}{8\pi}\int_{\frac m2}^{\infty}\frac{r^2(2-\psi_0)\langle K^2\rangle}{\psi_0^7}dr\ .
\ee
The Penrose inequality for the total ADM mass is satisfied 
up to the second order in the exterior curvature  if 
\be\la{32d}
P_I\geq \bar p^2\ ,
\ee
where $\bar p$ is the total momentum  given by
\be\la{32e}
p^i=\frac{1}{8\pi}\lim_{r\rightarrow\infty}\oint{r^2K^i_{\ r}d\sigma}
\ee
in the asymptotic  Cartesian coordinates.
If the sharp version of inequality  (\ref{32d}) is satisfied then the Penrose inequality is true for a sufficiently small exterior curvature. If it is saturated we have to investigate higher order corrections.

\section{Axially symmetric perturbations}
If $H=0$ the momentum constraint in flat space equipped with the spherical coordinates $r,\theta,\varphi$ yields \cite{tj} 
\begin{equation}
K_{\varphi\theta}=\frac{\omega _{,r}}{\sin{\theta}}\ ,\ \ K_{\varphi r}=-\frac{\omega _{,\theta}}{r^2\sin{\theta}}\ , \label{28}
\end{equation}
\be
(r^3K_{rr}\sin{\theta})_{,r}+(rK_{r\theta}\sin{\theta})_{,\theta}=0\ ,\label{1}
\ee
\be\label{7}
(K_{\theta\theta}\sin^2{\theta})_{,\theta}+(r^2K_{r\theta})_{,r}\sin^2{\theta}+r^2K_{rr}\sin{\theta}\cos{\theta}=0\ ,
\ee
where $\omega$ is an arbitrary function. The general solution of equations (\ref{1})-(\ref{7}) can be given explicitly in terms of one function \cite{cm,c}. However, this description is not useful for our goals.

Let us  consider regularity conditions which should be satisfied by $K_{ij}$. It follows from (\ref{28}) that at $\theta=0,\pi$ derivatives of $\omega$ should vanish sufficiently fast,
\be\la{omega_assym}
\omega_{,\theta}\sim \sin^3{\theta}\ ,\ \ \omega_{,r}\sim \sin^4{\theta}\ .
\ee
Hence
\be\la{28i}
\omega=f\sin^4{\theta}+J(\cos^3{\theta}-3\cos{\theta})+c\ ,
\ee
where $f$ is a smooth function of $r$ and $z=\cos{\theta}$, $J$ is a constant which plays a role of the ADM angular momentum and $c$ is a constant which can be omitted.
Moreover, components  $K_{rr}$ and $K_{r\theta}/\sin{\theta}$ should be  smooth and 
the following  condition  should be satisfied in a neighbourhood of  the symmetry axis
\be
K_{\theta\theta}d\theta^2+K_{\varphi \varphi}d\phi^2=f_1(d\theta^2+\sin^2{\theta}d\varphi^2)+f_2d(\cos{\theta})^2\ ,
\label{8b}
\ee
where $f_i$ are smooth functions.
Hence $K_{\theta\theta}$ should be smooth and
\be
K^{\theta}_{\ \theta}=K^{\varphi}_{\ \varphi}+f_3\sin^2{\theta}\ .\label{8}
\ee
Since $H=0$ there is 
\be\la{8a}
K^{\varphi}_{\ \varphi}=-K^{\theta}_{\ \theta}-K^r_{\ r}
\ee
and equation (\ref{8}) is equivalent to
\be
2K^{\theta}_{\ \theta}=-K^{r}_{\ r}+f_4\sin^2{\theta}\ .\label{8c}
\ee
If $K_{\theta\theta}$ is differentiable with respect to $z$ then  (\ref{8c}) follows from (\ref{7}).  Thus, in addition to  (\ref{28i}) it is sufficient to require that  $K_{rr}$, 
$K_{\theta\theta}$ and $K_{r\theta}/\sin{\theta}$ are smooth functions of $r$ and $z$.

Equation (\ref{1}) can be solved in terms of a potential $Q$, 
\be\label{4}
K_{r\theta}=\frac{Q_{,r}}{r\sin{\theta}}\ ,
\ee
\be\label{6}
K_{rr}=-\frac{Q_{,\theta}}{r^3\sin{\theta}}\ .
\ee
Let us introduce a function $F$ such that 
\be\la{11}
K_{\theta\theta}+\frac 12r^2K_{rr}=F_{,z}\ .
\ee
Then  (\ref{7}) yields an equation for $F$,
\be\label{12a}
\bigtriangleup_s F=(rQ_{,r})_{,r}+\frac {1}{2r} (1-z^2)Q_{,zz}\ ,
\ee
where 
\be
\bigtriangleup_s F=((1-z^2)F_{,z})_{,z}
\ee
is the spherical Laplacian of $F$. 

The regularity conditions imply that $Q$ is everywhere smooth and
\be
Q_{,r}\sim (z^2-1)\ .\label{3}
\ee
Integrating (\ref{3}) over $r$ yields 
\be
Q=Q'(z^2-1)+ L(z)\ ,\label{3b}
\ee
where functions $Q'$ and $L$ are smooth. Function $L$ can be written in the form $L=L'+az+a'$, where $a,a'$ are constants and $ L'$ vanishes on the symmetry axis. Function $ L'$ can be incorporated in  $Q'$ and constant $a'$ has no effect on $K_{ij}$.  Without a loss of generality
we can assume that
\be
Q=(z^2-1)q_{,z}+az\ ,\ \ a=\mathrm{const}\ ,\label{3c}
\ee
where $q$ is a smooth function of $r$ and $z$. Substituting (\ref{3c}) into (\ref{12a}) yields
\be\label{12}
\bigtriangleup_s F=(z^2-1)\p_z((rq_{,r})_{,r}+\frac {1}{2r} \bigtriangleup_s q)\ .
\ee
Equation (\ref{12}) is solvable with respect to $F$ if the integral of its r.h.s. over $z$ vanishes. Hence, $q$ should satisfy
\be
\int_{-1}^1{zqdz}= pr+\frac {\tilde p}{r}\ , \la{19c}
\ee
where  $p,\tilde p$ are constants. Formula (\ref{32e}) shows that parameter $p$ is the value of the ADM  momentum   directed along the symmetry axis.

Another regularity condition for $q$ is related to 
the asymptotical flatness condition (\ref{9a}). It reads
\be\la{3a}
|q_{,r}|<\mathrm{const}<\infty\ .
\ee
Note that under condition (\ref{3a}) function $F_{,z}$ will be also  bounded.

The first two terms in expression (\ref{32b}) depend exclusively on the restriction of function $K_{rr}$  to $r=\frac m2$. Given it
we would like to find the minimal value of the  last term  in (\ref{32b}).  This term has the form
\be\la{17}
\frac{1}{2\pi}\int_{\frac m2}^{\infty}{\varrho r^5\langle K^2\rangle dr}=P_J+\tilde P_I\ ,
\ee
where
\be\la{18}
\varrho=\frac {m(1-\frac {m}{2r})}{4r^3(1+\frac {m}{2r})^7}\ ,
\ee
\be\la{19a}
P_J=2\int_{\frac m2}^{\infty}{dr \varrho r\int_{-1}^1(K_{\varphi\theta}^2+r^2K_{\varphi r}^2 )\frac{dz}{1-z^2}}\ ,
\ee
\be\la{19}
\tilde P_I=\int_{\frac m2}^{\infty}{dr \varrho r\int_{-1}^1\big (2r^2K_{r\theta}^2+2(K_{\theta\theta}+\frac 12 r^2K_{rr})^2+\frac 32r^4K_{rr}^2\big )}dz
 \ .\ee
Since $\omega$ is not related to $K_{rr}$, the minimal value of $P_J$ is zero. In order to estimate minimum of $\tilde P_I$ 
let us decompose $q$ into the Legendre polynomials $P_n$ 
\be\la{21}
q=\Sigma_1^{\infty} q_nP_n\ ,
\ee
where
\be
q_1=\frac 32( pr+\frac {\tilde p}{r}) \la{19b}
\ee
(see (\ref{19c}) and note that $q_0=0$ can be assumed)
and other coefficients $q_n$ are unknown functions of $r$. Thanks to the standard property of the Legendre polynomials,
\be\la{17c}
\bigtriangleup_s P_n=-n(n+1)P_n\ ,
\ee
from (\ref{6}) one obtains
\be\la{23}
K_{rr}=\frac{1}{r^3}(a+\Sigma_1^{\infty} n(n+1) q_nP_n)
\ee
and
\be\la{23a}
\int_{-1}^{1}{K_{rr}^2}dz=\frac{2}{r^6}(a^2+\Sigma_1^{\infty} \frac{n^2(n+1)^2}{2n+1}(q_{n})^2)\ .
\ee
From (\ref{4}) it follows that
\be\la{22a}
\int_{-1}^{1}{K_{r\theta}^2dz}=\frac{1}{r^2}\int{q_{,r}((z^2-1)q_{,z})_{,zr}dz}=\frac{2}{r^2}\Sigma_1^{\infty} \frac{n(n+1)}{2n+1}(q_{n,r})^2\ .
\ee
Using again  (\ref{17c}) and another identity,
\be\la{17b}
(z^2-1)P_{n,z}=\frac{n(n+1)}{2n+1}(P_{n+1}-P_{n-1})\ ,
\ee 
one can easily find   a solution $F$ of (\ref{12}). Hence
\be\la{24b}
K_{\theta\theta}+\frac 12r^2K_{rr}=\Sigma_2^{\infty}\big ((rq_{n,r})_{,r}-\frac 12n(n+1)\frac{q_n}{r}\big)\tilde P_{n}\ ,
\ee
where
\be\la{34}
\tilde P_n=\p_z(\bigtriangleup_s^{-1}((z^2-1)\p_zP_n))=\frac{1}{n+2}(nP_n-\frac{2}{n-1}P_{n-1,z})\ ,\ n\geq 2\ .
\ee

Polynomials $\tilde P_n$ are orthogonal 
\be\la{35}
\int_{-1}^1{\tilde P_k\tilde P_ndz}=c_n\delta_{kn}\ ,
\ee
where
\be\la{36}
c_n=\frac{2n(n+1)}{(n-1)(n+2)(2n+1)}\ .
\ee
In order to show (\ref{35}) let us assume that $k\leq n$. Then
\be\la{28b}
\int_{-1}^1{P_{n-1,z}P_{k-1,z}dz}=(P_{n-1}P_{k-1,z})_{-1}^1
\ee
and
\be\la{28d}
\int_{-1}^1{P_{k-1,z}P_ndz}=0\ ,
\ee
since polynomial $P_{k-1,zz}$ is of lower order than $P_{n-1}$ and $P_{k-1,z}$ is of lower order than $P_n$. From   the standard identities
\be\la{28c}
P_{k,z}=P_{k-2,z}+(2k-1)P_{k-1}\ , 
\ee
\be\la{28e}
\int_{-1}^1{P_kP_ndz}=\frac{2}{2n+1}\delta_{kn}
\ee
one obtains
\be\la{28g}
\int_{-1}^1{P_{n-1,z}P_kdz}=(P_{n-1}P_k)_{-1}^1-\int{P_{n-1}P_{k,z}dz}=(P_{n-1}P_k)_{-1}^1-2\delta_{kn}\ .
\ee
Formulas (\ref{28b})-(\ref{28c}) lead to (\ref{35}).
Note that boundary terms present in (\ref{28b}) and (\ref{28g}) cancel each other  due to the properties
\be\la{26c}
P_{k-1,z}(\pm 1)=\frac 12(\pm 1)^kk(k-1)\ ,\ \ P_k(\pm 1)=(\pm 1)^k\ .
\ee

It follows from (\ref{24b}) and (\ref{35}) that
\be\la{37}
\int_{-1}^{1}(K_{\theta\theta}+\frac 12r^2K_{rr})^2dz=\Sigma_2^{\infty} c_n\big ((rq_{n,r})_{,r}-\frac 12n(n+1)\frac{q_n}{r}\big)^2\ .
\ee
Substituting (\ref{23a}), (\ref{22a}) and (\ref{37}), with $q_1$ given by (\ref{19b}), to (\ref{19}) yields
\be\la{38}
\tilde P_I=\Sigma_0^{\infty}I_n\ ,
\ee
where
\be\la{39}
I_0+I_1=3\int_{\frac m2}^{\infty}\frac{\varrho}{r}(a^2+2p\tilde p+5\frac{\tilde p^2}{r^2}+5p^2r^2)dr=\frac{a^2}{32 m^2}+\frac{p\tilde p}{16 m^2}+\frac{\tilde p^2}{8 m^4}+\frac{129 p^2}{128}\ ,
\ee
\be\la{40a}
I_{n\geq 2}=2c_n\int_{\ln{\frac m2}}^{\infty}\varrho  \big [(\ddot q_n-Nq_n)^2+(N-1)(2\dot q_n^2+3Nq_n^2)\big ]du
\ee
and symbol $N$ and new coordinate $u$ are defined by
\be
N=\frac 12n(n+1)\ ,\ \ r=e^u\ ,\ \ \dot q_n=\p_uq_n\ .
\ee
 Let $F_{n\geq 2}$ be a space of functions $q_n$ with fixed value $q_{nh}$ at $r=\frac m2$ and $y_n\in F_n$ be  a solution of equation following from  the variational principle for $I_n$,
\be\la{41}
(\p_u^2-N)[\varrho(\p_u^2-N)y_n]-2(N-1)\p_u(\varrho \p_u y_n)+3N(N-1)\varrho y_n=0\ .
\ee
We will show that $I_n(y_n)$ is the absolute minimum of $I_n$ on space $F_n$. To this end we write a general function from $F_n$ as $q_n=y_n+q'_n$, where $q'_n=0$ at $r=\frac m2$.  Then
\ba\la{42}
&I_n(q_n)=I_n(y_n)+I_n(q'_n)+\hspace{7cm}
\\\nonumber
&4c_n\int_{\ln{\frac m2}}^{\infty}\varrho  \big [(\ddot q'_n-Nq'_n)(\ddot y_n-Ny_n)+(N-1)(2\dot q'_n\dot y_n+3N q'_ny_n)\big ]du\ .
\ea
If we remove  derivatives of $y_n$  via integration by parts we see that the integral over $u$ in (\ref{42})  disappears thanks to (\ref{41}) and vanishing boundary values of $\varrho$ and $q'_n$. Since $I_n(q'_n)\geq 0$ we obtain
\be\la{43}
I_n(q_n)\geq I_n(y_n)
\ee
and this inequality is saturated only if $q_n=y_n$. This result also shows that solution of (\ref{41}), if it exists, is unique (otherwise two different solutions, say $y_n$ and $y'_n$, would have to satisfy $I_n(y'_n)> I_n(y_n)$ together with $I_n(y_n)> I_n(y'_n)$). Thus, solution of (\ref{41}) is defined exclusively by the boundary value $y_{nh}=q_{nh}$ at $r=\frac m2$. Value of $I_n(y_n)$ is also given by boundary terms,
\be\la{44}
I_n(y_n)=\frac{c_n}{32m^2}y_{nh}(\ddot y_n-Ny_n)_h\ ,
\ee
but, in order to find $\ddot y_n$ at $r=\frac m2$, we have first to solve (\ref{41}). Fortunately,  using the symbolic programme Mathematica and  method of trial end error we were able to find exact form of $y_n$,
\be\la{44a}
y_n = \frac{Ny_{nh}}{2N+1} \left( \frac{n-1}{n+1} \left( \frac{m}{2r}\right)^{n+1} + \frac{n+2}{n} \left( \frac{m}{2r}\right)^n\right).
\ee
It follows from (\ref{44}) and (\ref{44a}) that
\be\la{44b}
I_n(y_n)=\frac{n^2(n+1)^2}{32m^2(n^2+n+1)(2n+1)}y_{nh}^2\ .
\ee

Axially symmetric and asymptotically vanishing solution of equation (\ref{14a})  has the form
\be\la{45}
\psi_1=\Sigma_0^{\infty}a_n(\frac{m}{2r})^{n+1}P_n\ .
\ee
Coefficients $a_n$  follow from condition (\ref{15}),
\be\la{46}
a_0=-\frac{a}{4m^2}\ ,\ \ a_{n\geq 1}=-\frac{n(n+1)}{4m^2(2n+1)}q_{nh}\ .
\ee
Hence
\be\la{47}
\langle \psi_1^2\rangle_{h}=\frac{\pi}{4m^4}(a^2+\Sigma_1^{\infty}\frac{n^2(n+1)^2}{(2n+1)^3}q_{nh}^2)\ .
\ee

Taking into  account equations (\ref{17}),  (\ref{38}), (\ref{39}), (\ref{43}), (\ref{44b}) and (\ref{47}),  we obtain the following inequality for the quantity $P_I$ defined by (\ref{32b})
\be\la{48}
P_I-p^2\geq P_J+\frac{1}{32m^2}\Sigma_2^{\infty}\frac{n^2(n+1)^2(n-1)(n+2)}{(n^2+n+1)(2n+1)^3}q_{nh}^2.
\ee
Here $P_J$ is given by (\ref{19a}), so the r.h.s. of (\ref{48}) is always nonnegative. Inequality (\ref{48}) is saturated if $q_n=y_n$. 
If the angular momentum $J$ is nontrivial then $d\omega\neq 0$ and $P_J>0$. In order  to estimate a lower bound of $P_J$ in terms of $J$ let us write expression  (\ref{28i}) in the form
\be\la{50}
\omega=\tilde\omega+J(z^3-3z)+c\ ,\ \ \tilde\omega=f(1-z^2)^2\ .
\ee
Then
\be \la{51}
\begin{aligned}
& \int_{-1}^1\frac{\omega_{,z}^2}{1-z^2}dz=\int_{-1}^1\frac{\tilde\omega_{,z}^2}{1-z^2}dz-6J\int_{-1}^1\tilde\omega_{,z}dz+9J^2\int_{-1}^1(1-z^2)dz = \\
 =& \int_{-1}^1\frac{\tilde\omega_{,z}^2}{1-z^2}dz+12J^2\ .
\end{aligned}
\ee
Hence
\be\la{52}
P_J\geq \frac{J^2}{4m^2}
\ee
and this inequality is saturated if $f=0$. It is equivalent, up to $K^2$, to 
\be\la{53}
P_J\geq \frac{4 \pi}{|S_h|}J^2
\ee
taking into account the expansion
\be
\frac{4 \pi}{|S_h|}J^2 = \frac{J^2}{4 m^2} -\frac{J^2}{8 \pi m^2} \langle \psi_1 \rangle_h - \frac{J^2 }{16 \pi^2 m^2} \left( 2 \pi \bigg\langle \psi_2 + \frac{3}{4} \psi_1^2 \bigg \rangle_h -\langle \psi_1 \rangle_h^2 \right) + ...
\ee
Substituting (\ref{53}) into  (\ref{48})  written in terms of multipole moments of  $K_{rr}$ yields
\be\la{53a}
E^2-p^2\geq \frac{|S_h|}{16\pi}+\frac{4 \pi}{|S_h|}J^2+2^{-11}m^4\Sigma_2^{\infty}\frac{(n-1)(n+2)}{(n^2+n+1)(2n+1)^3}(K_{rr})_{h}^2.
\ee
Inequality (\ref{53a})  is saturated if $f=0$ and $q_n=y_n$. If $\omega\neq J(z^3-3z)+c$ or $q_n\neq y_n$ or one of the moments of $(K_{rr})_h$ with $n\geq 2$ does not vanish then  the sharp form of the strong Penrose inequality (\ref{8k}) is necessarily satisfied in the order $K^2$.  

\section{Generalized Bowen-York initial data}
Inequality (\ref{53a})  does not lead to  (\ref{8k})  only if  $\omega=J(z^3-3z)+c$ and $q=q_1z$. In this case tensor $K_{ij}$ is given  by  
\be\la{49}
 K_{rr}=\frac {a}{r^3}+3(\frac {\tilde p}{r^4}+\frac{p}{r^2})\cos{\theta}\ ,\ \ K^{\theta}_{\ \theta}=K^{\varphi}_{\ \varphi}=-\frac 12K_{rr}\ ,
\ee
\ba\nonumber
 K_{r\theta}=\frac 32(\frac {\tilde p}{r^3}-\frac pr)\sin{\theta}\ ,\ \ K_{\varphi r}= \frac{3J}{r^2}(z^2-1)\ , \ \      K_{\varphi\theta}=0\ ,
\ea
where  $a,p,\tilde p,J$ are constants. This form of $K_{ij}$ is equivalent to 
 the generalized Bowen-York extrinsic curvature \cite{beig}. Parameters $p$ and $\tilde p/m^2$ are the total momenta viewed, respectively, from $r=\infty$ and $r=0$ (the latter point can be thought as another infinity obtained by an inversion). The constant $a$ does not have any obvious physical interpretation.

In order to check the Penrose inequality in the case  (\ref{49}) we have  to consider terms in the energy (\ref{13}) and surface area (\ref{26}) of the third order in $K$. These take the following form
\be \la{e3}
E^{(3)} = - \frac{7}{16 \pi} \int_{\frac m2}^{\infty}\frac{r^2}{\psi_0^7} \langle K^2 \psi_1 \rangle dr + \frac{ 3 m^2}{256 \pi} \langle K_{rr} \left( \psi_2 - \psi_1^2 \right) \rangle_h\ ,
\ee
\be \la{s3}
|S_h|^{(3)} = 16 m^2 \bigg\langle \frac{\psi_1^3}{8} + \frac{3}{4} \psi_1 \psi_2 + \tilde \psi_3  \bigg\rangle_h.
\ee
Following derivation of (\ref{29}) one obtains
\be
\langle \tilde \psi_3 \rangle_h = - \frac{3m}{128} \langle K_{rr} \left( \psi_1^2 -\psi_2 \right) \rangle_h -\frac{7}{8} \int_{m/2}^{\infty}{\frac{dr'}{r'^2\psi_0^2}\int_{\frac m2}^{r'}{\frac{\tilde r^2}{\psi_0^7} \langle K^2 \psi_1 \rangle d\tilde r}}.
\ee
The third order correction to  (\ref{32b})  reads
\be \la{P3}
\begin{aligned}
P_I^{(3)}= & - \frac{7m}{8 \pi} \int_{m/2}^{\infty} \frac{r^2 (2-\psi_0) \langle K^{2} \psi_1 \rangle }{\psi_0^8}  \mathrm{d}r  - \frac{m^2 \langle K_{rr} \rangle_h }{1024\pi^2} \int_{m/2}^{\infty} \frac{r^2 \langle K^2 \rangle}{\psi_0^6} \mathrm{d}r -  \\
& - \frac{m^2}{4 \pi} \bigg\langle \frac{1}{2}\psi_1^3 + 3 \psi_1 \psi_2 \bigg\rangle_h - \frac{3m^4}{16384 \pi^2} \langle K_{rr} \psi_1 \rangle_h \langle K_{rr}\rangle_h.
\end{aligned}
\ee
To compute it  we  need an explicit formula for a solution $\psi_2$ of the  Poisson equation
\be \la{psi2}
\Delta \psi_2 = -\frac{1}{8 \psi_0^7} K^2
\ee
 with the boundary condition
 \be \la{psi2b}
\partial_r \psi_2 +  \frac{1}{m} \psi_2 = -\frac{3}{64} K_{rr} \psi_1 \ \ \mathrm{at} \ \ r=m/2.
\ee
For data (\ref{49}) we can expand both sides of the above equations  into the Legendre polynomials. Only few first coefficients in $\psi_2$ survive and they can be found explicitly using the symbolic programme Mathematica. Surprisingly, expression $(P_I-\frac{4 \pi}{|S_h|}J^2)^{(3)}$ vanishes (we suppose that cancellation of terms in this expression has some deeper reasons, not  known to us) and we have to pass to the fourth order calculus.

Then in place of  (\ref{e3})-(\ref{P3}) one obtains 
\be \la{e4}
E^{(4)} = \frac{7}{16 \pi} \int_{\frac m2}^{\infty}\frac{r^2}{\psi_0^7} \bigg\langle K^2  \left(4 \frac{\psi_1^2}{\psi_0} -  \psi_2 \right) \bigg\rangle dr - \frac{ m^2}{512 \pi} \langle K_{rr} \left( 12 \psi_1 \psi_2 - 6 \psi_3 - 5\psi_1^3 \right) \rangle_h\ ,
\ee
\be \la{s4}
|S_h|^{(4)} = 4 m^2 \bigg\langle \frac{\psi_1^4}{16} +\frac{3 \psi_1^2 \psi_2}{2} + \frac{3 \psi_2^2}{2} + 3 \psi_1 \psi_3 + 4 \tilde \psi_4  \bigg\rangle_h\ ,
\ee
\be
\begin{aligned}
& \langle \tilde \psi_4 \rangle   = -\frac{m}{256} \left< K_{rr} \left(12 \psi_1 \psi_2 - 6 \psi_3 - 5\psi_1^3 \right)\right>\Bigr|_h  \\
& + \frac{7}{8} \int_{m/2}^{\infty}\frac{\mathrm{d}r'}{r^{'2}\psi_0^2}\int_{\frac{m}{2}}^{r'} \frac{\tilde r^2}{\psi_0^7} \bigg\langle K^2  \left(4 \frac{\psi_1^2}{\psi_0} -  \psi_2 \right) \bigg\rangle \mathrm{d} \tilde r \ ,
\end{aligned}
\ee
\be \la{p4}
\begin{aligned}
P_I^{(4)} & = \frac{1}{256 \pi^2} \left(\int_{m/2}^{\infty} \frac{r^2 \langle K^2 \rangle }{\psi_0^6}  \mathrm{d}r \right)^2 + \frac{9m^4}{65536 \pi^2} \langle K_{rr}\psi_1\rangle_h^2  
\\
& - \frac{m^2}{16 \pi} \bigg\langle \frac{\psi_1^4}{4} + 6 \psi_1^2 \psi_2 + 6 \psi_2^2 + 12 \psi_1 \psi_3 \bigg \rangle_h -  \frac{3m^4}{16384 \pi^2}\langle  K_{rr}\left( \psi_2 - \psi_1^2\right)\rangle_h \langle K_{rr} \rangle_h  \\
& + \frac{7 m}{8 \pi} \int_{m/2}^{\infty}  \frac{r^2 (2-\psi_0)}{\psi_0^8} \bigg\langle K^2 \left(  4 \frac{\psi_1^2}{\psi_0} - \psi_2 \right) \bigg\rangle \mathrm{d}r  \\
& + \frac{3m^2 \langle K_{rr}\psi_1\rangle_h }{2048 \pi^2} \int_{m/2}^{\infty} \frac{r^2 \langle K^2 \rangle }{\psi_0^6}  \mathrm{d}r + \frac{7m^2 \langle K_{rr}\rangle_h }{1024 \pi^2} \int_h^{\infty} \frac{r^2 \langle K^2 \psi_1 \rangle}{\psi_0^7} \mathrm{d}r\ .
\end{aligned}
\ee
To compute $P_I^{(4)}$ we first solve (using Mathematica) the  Poisson equation for $\psi_3$
\be \la{psi3}
\Delta \psi_3 = \frac{7}{8 \psi_0^8 } K^2 \psi_1
\ee
with the boundary condition
\be
\partial_r \psi_3 +\frac{1}{m}\psi_3 = \frac{3}{64}K_{rr}\left(\psi^2_1 -\psi_2 \right) \ \mathrm{at} \ r=m/2.
\ee
Correction $(P_I-p^2-\frac{4 \pi}{|S_h|}J^2)^{(4)}$ for data (\ref{49}) takes the form of a fourth order homogeneous polynomial $W(p/m,\tilde p/m^3, J/m^2)$  independent of $a$
\be\la{p4a}
(P_I-p^2-\frac{4 \pi}{|S_h|}J^2)^{(4)}=m^2W=0,088\frac{p^4}{m^2}+...
\ee
 If $p\neq 0$ then the term  $p^4$ dominates in $W$ and we have
 \be\la{p4b}
W>0,086\frac{p^4}{m^4}\ .
\ee
 If $p=0$ then
\be\la{p4c}
W\geq  0,004\frac{\tilde p^4}{m^{12}}+0,044\frac{\tilde p^2J^2}{m^{10}}+0,013\frac{J^4}{m^{8}}\ .
\ee
 Thus, if any of the parameters $p,\tilde p,J$ is nontrivial then the sharp Penrose inequality (\ref{8k}) is satisfied in the fourth order in $K$. If $p=\tilde p=J=0$ then data (\ref{49}) are spherically symmetric, so they should correspond to the Schwarzschild solution. Indeed, this metric admits sections of the form $t=f(r)$ which are conformally equivalent to (\ref{49}) \cite{mm}. Thus, in this case  the Penrose inequality is saturated.

 \section{Summary and discussion}
 Summarizing the  last two sections we can formulate our main result  in the following way.
 
 \begin{thm}
  Let $S=R^3\backslash B(0,\frac m2)$ be an initial surface bounded by the sphere $S_h$ with radius $m/2$. Let  $g_{ij}$ be flat metric on $S$ and $K_{ij}$ be a traceless axially symmetric solution of the momentum constraint satisfying condition (\ref{9a}). Assume that
 \begin{itemize}
 \item 
 the Lichnerowicz equation (\ref{7a}) admits solution $\psi>0$ satisfying the boundary condition (\ref{8f}) on  $S_h$  and the asymptotic condition $\psi\longrightarrow 1$ 
 \item
 $\psi$ can be expanded into powers of a parameter $\epsilon$ proportional to a norm of tensor $K_{ij}$.
 \end{itemize}
 Then  initial data (\ref{6a}) satisfy the Penrose inequality (\ref{8k}) up to the second order in $\epsilon$ in generic case and up to the fourth order in the case of generalized Bowen-York data (\ref{49}). Inequality (\ref{8k}) is saturated only in the case (\ref{49}) with\\ $p=\tilde p=J=0$ which corresponds to the Schwarzschild metric.
 \end{thm}
 Note that the conformal factor $\psi>0$ is known to exist for $K_{rr}$ of definte sign on $S_h$ (see discussion in the begining of Section 2).
 A more serious  problem is lack of  a criterion of smallness of $K_{ij}$. 
 Since $K^2$ is integrable we can write $K_{ij}=\epsilon \hat K_{ij}$, where the integral of $\hat K^2$ is 1. If $\psi$ exists and is differentiable with respect to $\epsilon$ we can expand it into the Taylor series in $\epsilon$. Our results  correspond to the leading terms ($\epsilon^2$ or $\epsilon^4$) in an expansion of (\ref{8k}). Thus, the Penrose inequality is satisfied for sufficiently small $\epsilon$, but it is difficult to estimate a range of $\epsilon$.
 In order to have any idea when our approximation agrees with exact results we solved the Lichnerowicz equation numerically in the case  (\ref{49}) with $a=\tilde p=0$ using FreeFEM solver \cite{freefem}.
We followed an approach in \cite{karkowski}, where  $r, \theta$ were replaced by coordinates $x=\frac{r}{r+m/2}$  and $y=\cos \theta$ which cover the finite domain $[1/2,1] \times [-1,1]$. Numerical results have been obtained for the specific choice of Schwarzschild mass ($m=2$) and then generalized to any value of $m$ by a simple rescaling of variables. Their comparison with our perturbative results is shown in
 Figure 1, where the intensity of the black color corresponds to the difference between  expression (\ref{p4a}) and numerical  value of $P_I - p^2 - \frac{J^2}{4 \pi |S_h|}$ which is positive in all considered cases. Since this function is independent  of sign of $p$ and $J$ it is sufficient to consider only positive values of these parameters. The plot has been obtained by approximating between 273 grid points spaced equidistantly in the domain $[0,1.7] \times [0,3]$ of variables $p/m$ and $J/m^2$.  It can be seen that, roughly, for  $|p|<0,5m$ and $|J|<2,5m^2$  perturbative formula (\ref{p4a}) underestimates a real value of $P_I - p^2 - \frac{J^2}{4 \pi |S_h|}$. This situation changes for higher values of $|p|$ and $|J|$.

\begin{figure}[h!]
\centering
\includegraphics[scale=0.308]{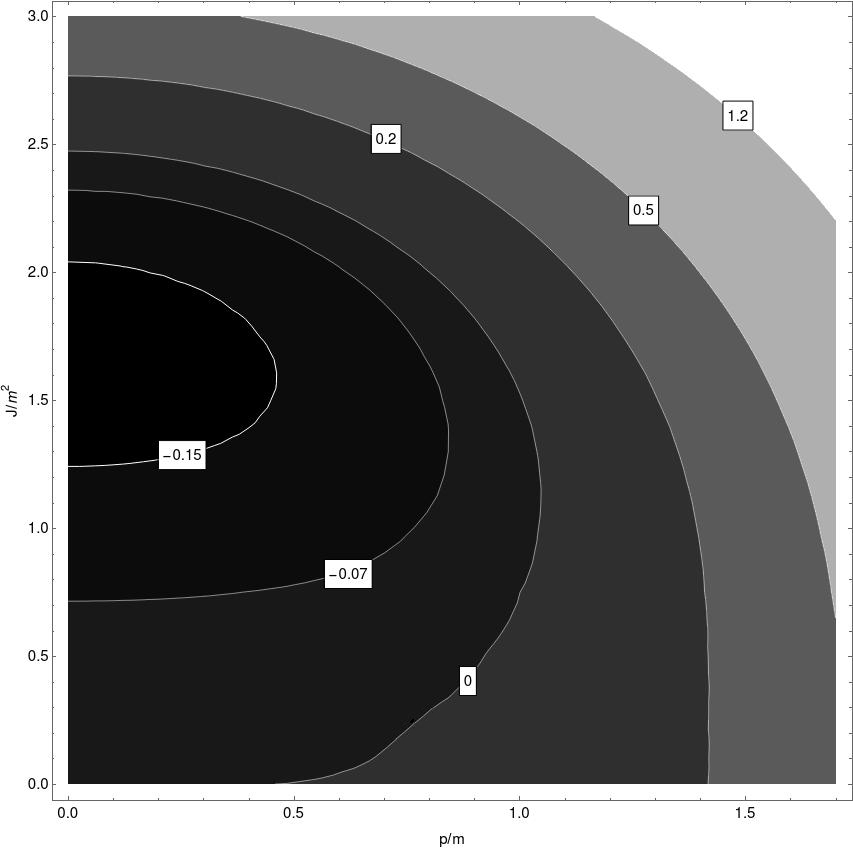}
\captionof{figure}{The difference between approximate (up to the fourth order of expansion in $\epsilon$) and numerical values of $\frac{1}{m^{2}}(P_I - p^2 - \frac{J^2}{4 \pi |S_h|})$.}
\end{figure}

Problems with fixing admitted value of the parameter $\epsilon$ is  a disadvantage of our approach when compared to the Geroch method. On the other hand the latter method seems suitable only for  the Penrose inequality  in the case when the inner boundary is an outermost minimal surface (see \cite{mms,rm} for an exception). 

Our results concern a class of initial data which depend at most on two free functions of two variables. In order to generalize them within the set of conformally flat data one should admit nonmaximal data ($H\neq 0$) depending on the azimuthal angle $\varphi$ and horizons which are not spheres. We continue our research in this direction.

\end{document}